\documentclass[superscriptaddress,prl,floatfix,twocolumn,amsmath,amssymb,aps]{revtex4-2}

\usepackage{graphicx}
\usepackage{hyperref}
\usepackage{amsfonts}
\usepackage{color}

\begin{document}
\title{Three New Phases in the Two-Dimensional Potts Model at Complex Temperature}
\author{Fang-Cheng Wang}
\affiliation{State Key Laboratory for Artificial Microstructure and Mesoscopic Physics, Frontier Science Center for Nano-optoelectronics, School of Physics, Peking University, Beijing 100871, People's Republic of China}
\author{Qi-Jun Ye}
\email{qjye@pku.edu.cn}
\affiliation{State Key Laboratory for Artificial Microstructure and Mesoscopic Physics, Frontier Science Center for Nano-optoelectronics, School of Physics, Peking University, Beijing 100871, People's Republic of China}
\affiliation{Interdisciplinary Institute of Light-Element Quantum Materials, Research Center for Light-Element Advanced Materials, and Collaborative Innovation Center of Quantum Matter, Peking University, Beijing 100871, People's Republic of China}
\author{Xin-Zheng Li}
\email{xzli@pku.edu.cn}
\affiliation{State Key Laboratory for Artificial Microstructure and Mesoscopic Physics, Frontier Science Center for Nano-optoelectronics, School of Physics, Peking University, Beijing 100871, People's Republic of China}
\affiliation{Interdisciplinary Institute of Light-Element Quantum Materials, Research Center for Light-Element Advanced Materials, and Collaborative Innovation Center of Quantum Matter, Peking University, Beijing 100871, People's Republic of China}
\affiliation{Peking University Yangtze Delta Institute of Optoelectronics, Nantong, Jiangsu 226010, People's Republic of China}
\date{\today}

\begin{abstract}
	The complex-temperature phase structure of the two-dimensional Potts model has remained largely unresolved despite decades of study.
	Here we determine the complex-temperature phase diagram of the 3-state Potts model, and uncover three gapless phases beyond the familiar ferromagnetic and paramagnetic phases.
	This is done by combining tensor-network free energies with the Poisson equation to compute the Fisher zero density in the thermodynamic limit.
	The resulting zero distribution unifies and revises long-standing conjectures on the zero loci, and outlines each phase.
	One is a non-conformal gapless phase, with an exponentially growing number of low-lying transfer-matrix levels distributed across all momentum sectors.
	The other two are Luttinger-liquid phases with complex-valued Luttinger parameters related by the Kramers--Wannier duality.
	At the antiferromagnetic critical point, the Luttinger parameter equals 3, while its limit from the adjacent Luttinger-liquid phase is 6.
	These findings reveal rich phase structures and unconventional critical behavior in a classical textbook model.
\end{abstract}

\maketitle
\textcolor{blue}{\textit{Introduction}.}---The two-dimensional Potts model has been a versatile cornerstone of statistical mechanics~\cite{wu1982potts}.
It provides a unified setting for continuous and first-order phase transitions~\cite{baxter1973potts} and a testing ground of various theories and algorithms, connecting to gas adsorption~\cite{alexander1975adsorb,berker1978adsorb,domany1978adsorb}, percolation and graph coloring~\cite{sokal2001bounds}.
Beyond its well-established real-temperature criticality, decades of study have revealed additional structures in the complex-temperature regime, including the antiferromagnetic (AFM) critical point~\footnote{The AFM critical point and its dual are $z=-1\pm\sqrt{4-q}$. The real-temperature regime is $z\ge 0$. Hence, the AFM critical temperature is complex for $q>3$, while its dual temperature is complex for all $q$.} and its dual point~\cite{baxter1982critical,jacobsen2006antiferromagnetic}, the Berker--Kadanoff (BK) fixed point~\cite{saleur1991antiferromagnetic}, and critical points described by complex conformal field theories (CFTs)~\cite{gorbenko2018walking,gorbenko2018walking-ii,jacobsen2024lattice,tang2024reclaiming}.
These findings suggest a rich phase structure at complex temperature.
Yet even for the 3-state model without external field, the complex-temperature phase diagram is still unsettled~\cite{jacobsen2006complex,martin2019zeros}.
Whether a third phase exists beyond the ferromagnetic (FM) and paramagnetic (PM) ones has remained an open question.
The Yang--Lee theory provides a universal framework to study systems with complex variables~\cite{yang1952statistical,lee1952statistical,fisher1965nature}.
It relates phases and phase transitions to the partition-function zeros on the complex plane, which typically coalesce into curves that delineate phase boundaries in the thermodynamic limit.
These zeros can be observed experimentally~\cite{peng2015experimental,gao2024experimental,laba2025quantum,kuzmak2025detecting}.
Their distribution not only encodes transition orders and critical exponents~\cite{bena2005statistical,abe1967note,kenna2006scaling}, but also connects to dynamical quantum phase transitions~\cite{heyl2013dynamical}, quantum many-body scars~\cite{meng2025detecting}, and the fermion sign problem~\cite{he2026fermion}.
However, computing these zeros for generic interacting models remains challenging, with exact calculations often limited to small lattices~\cite{pearson1982partition,kim2004yang,kim2004density,martin2019zeros}.
For Fisher zeros, i.e., partition function zeros on the complex-temperature plane, this challenge is further compounded, since they often form exotic topologies and even dense areas in the absence of the Lee--Yang circle theorem~\cite{lv2026giant,van1984location,derrida1991zeroes,janke2004phase}.
In this field, the square-lattice Potts model is a paradigmatic system, whose intricate Fisher zero distribution has attracted persisting interest and motivated numerous conjectures~\cite{chen1996partition,matveev1996complex,kim2004density,jacobsen2006complex,martin2019zeros}.
However, strong finite-size effects obscure which structures will survive in the thermodynamic limit~\cite{martin2019zeros}.
\begin{figure*}[t!]
  \centering
  \includegraphics[width=\textwidth]{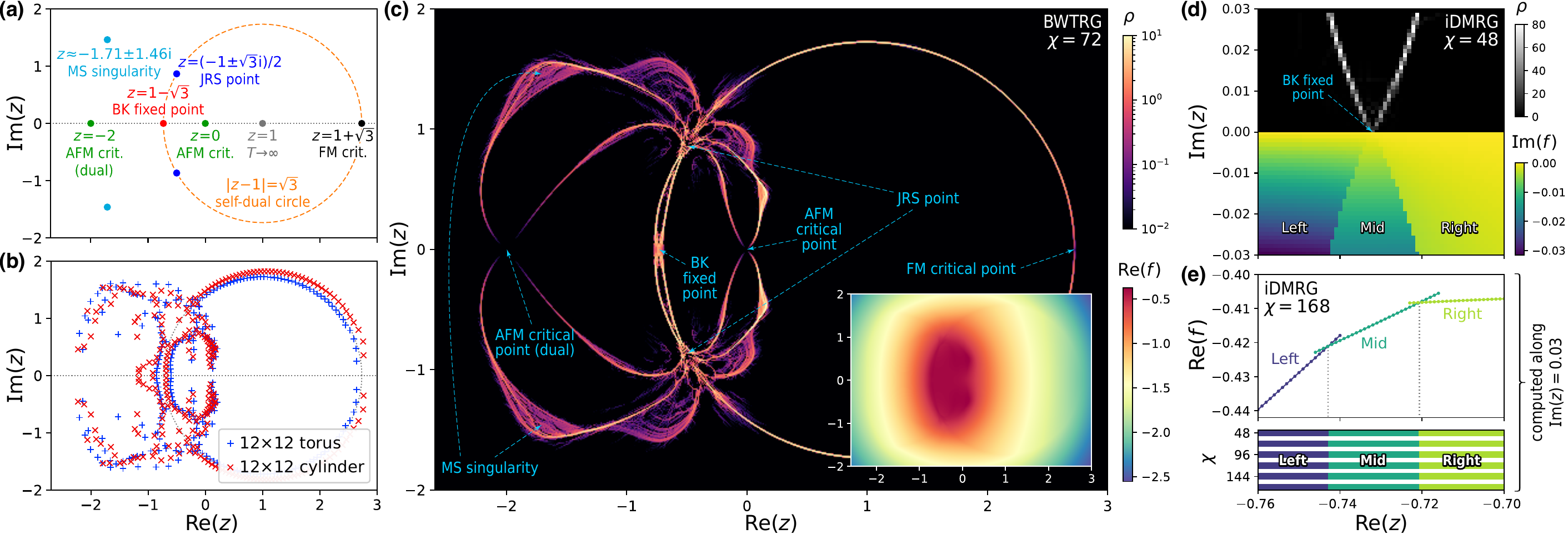}
  \caption{(a) Complex $z$ points relevant to established results and conjectures. (b) Fisher zeros obtained from finite-size partition functions. (c) Zero density $\rho(z)$ computed by BWTRG. Regions with $\rho(z)<10^{-2}$ are colored in black for simplicity. The inset shows the $\operatorname{Re}f(z)$ landscape. (d) $\rho(z)$ and $\operatorname{Im}f(z)$ near $z_\text{BK}$ computed by iDMRG. (e) Free-energy competition among the three phases near $z_\text{BK}$ along $\operatorname{Im}(z)=0.03$. Phase boundaries remain almost unchanged as $\chi$ increases.}
  \label{fig1}
\end{figure*}
In this Letter, we combine tensor-network free energies with the Poisson equation to compute the global Fisher zero density of the 3-state Potts model.
This approach works directly in the thermodynamic limit and requires no \textit{a priori} knowledge of the zero loci.
The resulting zero distribution unifies and revises long-standing conjectures, identifies the BK fixed point as a triple point, and reveals a complex-temperature phase diagram featuring three gapless phases.
One is non-conformal, with a number of low-lying transfer-matrix levels that grows exponentially with $L$ in every momentum sector.
The other two belong to the Luttinger-liquid (LL) universality class described by $c=1$ compact-boson CFT.
Their complex-valued Luttinger parameters and phase-diagram regions are related by the Kramers--Wannier (KW) duality.
We also find that at the AFM critical point, an imaginary thermal perturbation drives the renormalization-group (RG) flow away from the LL fixed line and back to it at a distant fixed point, yielding a sharp discontinuity in the Luttinger parameter.
\textcolor{blue}{\textit{Model and Methods}.}---The energy of the classical Potts model is defined as $E(\{\sigma_j\}) = -J \sum_{\langle jk \rangle} \delta_{\sigma_j\sigma_k}$, where $\sigma_j \in \{1, \dots, q\}$ is the integer-valued spin.
When $q=2$, it reduces to the Ising model.
Here, we focus on the $q=3$ case on the square lattice.
Using the conventional variable $z \equiv e^{\beta J}$, the partition function
\begin{equation}\label{partition}
	Z(z)=\sum_{\{\sigma_j\}}\mathrm{e}^{-\beta E(\{\sigma_j\})}=\sum_{\{\sigma_j\}}\prod_{\langle jk \rangle }z^{\delta_{\sigma_j\sigma_k}}
\end{equation}
is a polynomial of $z$.
The number of its zeros on the complex $z$-plane is proportional to the number of lattice sites $N$.
In the thermodynamic limit, the normalized zero density $\rho(z)$ satisfies the Poisson equation~\cite{derrida1991zeroes,blythe2002lee,ouyang2024complex,abdelshafy2025yang}
\begin{equation}\label{density}
  \rho(z)=-\frac{1}{2\pi}\nabla^2 \operatorname{Re}f(z)
,\end{equation}
where $f(z)\equiv -\lim_{N\to\infty}\frac{1}{N} \ln Z(z)$ is the dimensionless free energy per site.
The KW duality implies that $f(z^\star) = f(z) + 2 \ln(z-1) - \ln 3$, where $z^\star \equiv 1 + 3/(z-1)$ is the KW dual of $z$~\cite{wu1982potts}.
The zero density thus satisfies $\rho(z^\star) = \lvert z-1 \rvert^4\rho(z) / 9$, which we exploit to identify and suppress numerical artifacts~\footnote{See Supplemental Material, at \textcolor{red}{http://xxx}, for details and additional discussions, where Refs.~\citep{chan2026extracting,xie2012coarse} are also cited.\label{sm}}.
Tensor network methods provide a powerful route to access complex partition functions in the thermodynamic limit~\cite{denbleyker2014controlling, garcia2015density}.
Here, we use a synergistic framework of four complementary algorithms.
The bond-weighted tensor renormalization group (BWTRG)~\cite{adachi2022bond} is adopted to compute $\operatorname{Re}f(z)$ on a dense $z$-grid, yielding $\rho(z)$ via the discrete Laplacian.
We also use the infinite density matrix renormalization group (iDMRG)~\cite{white1992density,nishino1995density,mcculloch2008infinite} to compare free energies of different phases.
To characterize each phase, we use the Loop-TNR algorithm~\cite{yang2017loop} to extract CFT data~\cite{blote1986conformal,affleck1988universal,gu2009tensor} and the exact diagonalization (ED) method to compute transfer-matrix spectra at finite lattice widths.
All numerical implementations are based on the Julia \texttt{QuantumKitHub} ecosystem~\cite{devos2025tensorkit,vanthilt2026practical,tnrkit,mpskit,krylovkit}.
\textcolor{blue}{\textit{Fisher Zeros}.}---Baxter's exact results have identified the FM critical point $z_\text{FM}=1+\sqrt{3}$, the AFM critical point $z_\text{AFM}=0$, and its KW dual $z_\text{AFM}^\star = -2$~\cite{baxter1973potts,baxter1982critical}.
According to the Yang--Lee theory, they must be accumulation points of zeros.
Chen, Hu, and Wu conjectured that all zeros with $\operatorname{Re}(z)>1$ lie on the self-dual circle $\lvert z-1\rvert =\sqrt{3}$ in the thermodynamic limit, which is strongly supported by finite-size calculations~\cite{chen1996partition}.
In the remaining region, the zero distribution is elusive.
Jacobsen, Richard, and Salas (JRS) conjectured that zero curves pass through or close to the BK fixed point $z_\text{BK}\equiv 1 - \sqrt{3} $, and pass through $z_\text{JRS}\equiv(-1\pm \sqrt{3}\mathrm{i}) / 2$~\cite{jacobsen2006complex}.
Matveev and Shrock (MS) identified a singularity at $z_\text{MS}\approx -1.71\pm 1.46\mathrm{i}$ via the low-temperature series expansion, and conjectured from finite-size zeros that $z_\text{MS}$ is an endpoint of a zero arc~\cite{matveev1996complex}.
These $z$ points are summarized in Fig.~\ref{fig1}(a), with Fisher zeros for $12\times 12$ lattices in Fig.~\ref{fig1}(b) for comparison.
One cannot infer how many zero curves will intersect the real $z$-axis, and whether the zeros will coalesce into curves or areas.
To directly compute the zero distribution in the thermodynamic limit, we use BWTRG with bond dimension $\chi=72$.
The resulting $\rho(z)$ reproduces the self-dual circle, and exhibits a ``butterfly'' pattern on the left [Fig.~\ref{fig1}(c)].
Three pairs of zero curves intersect at $z_\text{BK}$, $z_\text{AFM}$, and $z_\text{AFM}^\star$, respectively, forming three X-shaped structures.
Away from the real $z$-axis, they converge at $z_\text{JRS}$.
These features are robust under different $\chi$ and algorithms~\footref{sm}, strongly supporting the JRS conjectures on $z_\text{BK}$ and $z_\text{JRS}$.
While the zeros coalesce into curves, they also form areal density on the ``wing tips'' of the butterfly.
Note that $z_\text{MS}$ lies in the interior of a dense-zero area, rather than the previously conjectured endpoint.
Why the paradox?
Our explanation is that the series expansion only describes the analytic behavior of a \textit{single} phase, but zero curves are generally determined by two or more adjacent phases~\cite{biskup2000general}.
A singularity of the continued phase may occur where the phase is metastable.
In such a case, it need not appear in $f(z)$ and $\rho(z)$.
Intuitively, one may doubt that the two zero curves connecting $z_\text{BK}$ and $z_\text{JRS}$ seem too close.
Do they truly sandwich a phase, or will they merge into a single curve on the self-dual circle as $\chi\to\infty$?
To answer this question, we compute $f(z)$ near $z_\text{BK}$ using iDMRG.
The resulting $\rho(z)$ and $\operatorname{Im}f(z)$ show that $z_\text{BK}$ is indeed a triple point [Fig.~\ref{fig1}(d)].
We further compare the three competing free-energy branches from $z=-0.76+0.03\mathrm{i}$ to $-0.70+0.03\mathrm{i}$ with increasing $\chi$ [Fig.~\ref{fig1}(e)].
At each $z$, the branch with the smallest $\operatorname{Re}f(z)$ is stable and dominates the partition function~\cite{biskup2000general}.
Our results show that the sandwiched phase remains stable in a finite interval, which does not narrow as $\chi$ increases.
\textcolor{blue}{\textit{Phase Diagram}.}---The zero curves and areas partition the complex $z$-plane into zero-free regions.
Each of them represents a phase, in which $f(z)$ varies analytically.
For instance, the two regions containing $z_\text{FM}^-$ and $z_\text{FM}^+$ represent the well-studied PM and FM phases, respectively.
Next, we will identify two gapped phases, a non-conformal gapless (NCG) phase, and two conformal gapless phases of the LL universality class, as summarized in Fig.~\ref{fig2}(a).
Here ``gap'' refers to the radial structure of the transfer-matrix spectrum $\{\lambda_\alpha\}$.
We say the system is gapless if infinitely many eigenvalues satisfy $\ln \lvert \lambda_0 / \lambda_\alpha \rvert \to 0$ as $L\to\infty$, where $\lambda_0$ denotes the dominant eigenvalue.
It is useful to define the ``level'', i.e., the \textit{rescaled} transfer-matrix gap
\begin{equation}
  D_\alpha \equiv \frac{L}{2\pi}\ln \left\lvert\frac{\lambda_0}{\lambda_\alpha}\right\rvert
.\end{equation}
In a gapped system, a finite number of $D_\alpha\to 0$ while others go to infinity.
In a system described by a CFT, low-lying levels satisfy $D_\alpha\to\operatorname{Re}(\Delta_\alpha - \Delta_0)$, where $\Delta_\alpha$ is the operator scaling dimension~\cite{cardy1984conformal,cardy2010conformal}.
\begin{figure}[t!]
  \centering
  \includegraphics[width=\linewidth]{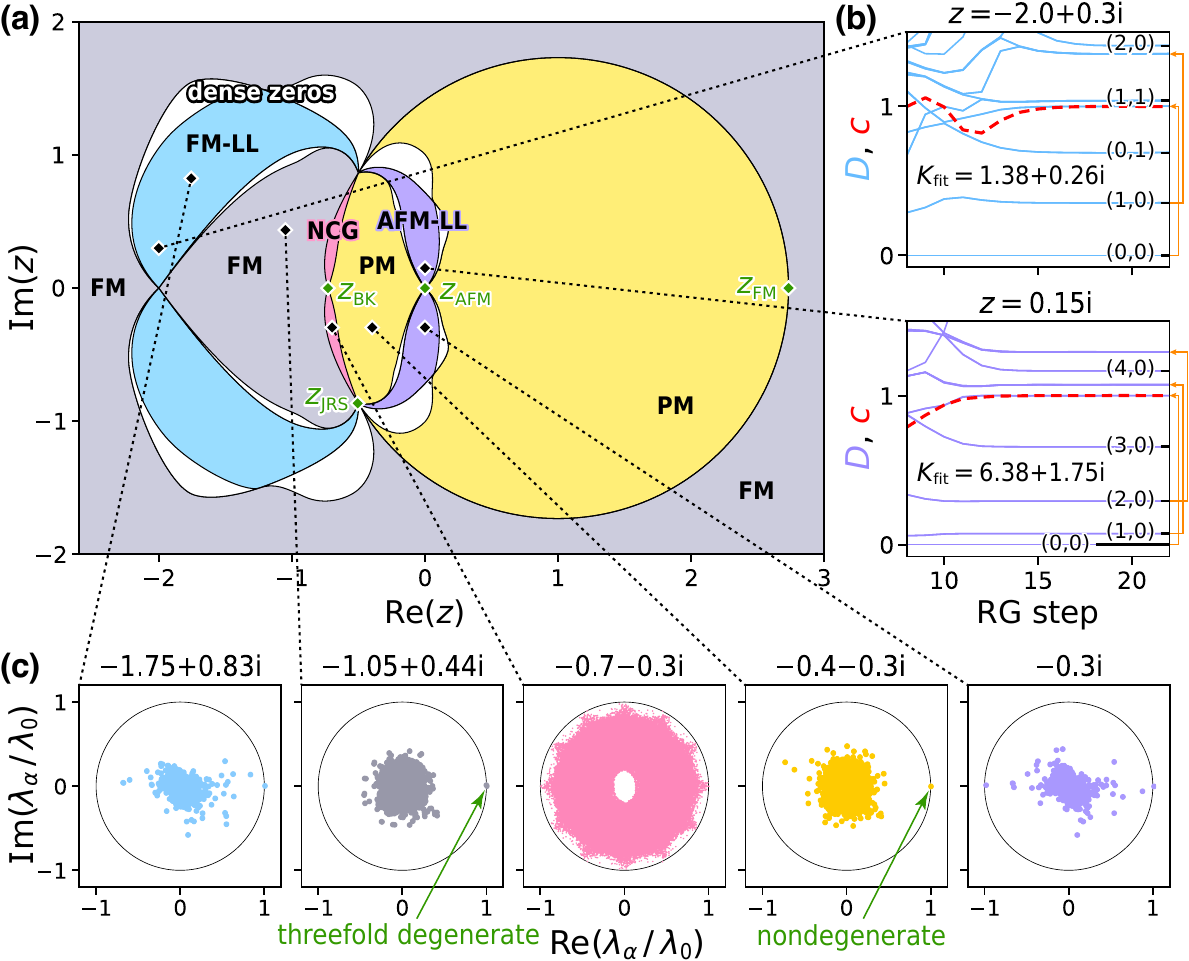}
  \caption{(a) Complex-temperature phase diagram of the 3-state Potts model. (b) Transfer-matrix levels $D$, effective central charges $c$, and Luttinger parameters $K_\text{fit}$ computed via Loop-TNR. Real parts of Eq.~(\ref{delta}) are labeled as $(n,m)$, whose descendants are indicated by orange arrows. (c) Transfer-matrix spectrum $\{\lambda_\alpha\}$ computed by ED ($L=12$) at representative $z$ points. Black solid lines denote the unit circle. The NCG phase exhibits a ``snowflake'' spectrum distinct from both the gapped phases and the LL phases.}
  \label{fig2}
\end{figure}
We run the Loop-TNR algorithm with $\chi=30$ at different $z$ points.
In the yellow (gray) regions of Fig.~\ref{fig2}(a), Loop-TNR converges to the high (low)-temperature trivial fixed point, where all transfer-matrix eigenvalues collapse to 0 except for a nondegenerate (threefold degenerate) $\lambda_0$, signifying a PM (FM) gapped phase.
In the cyan and purple regions, the RG flow is attracted to nontrivial fixed points, where the low-lying $\{\ln(\lambda_0 / \lambda_\alpha)\}$ exhibit integer ratios and descendant structures [Figs.~\ref{fig2}(b) and S1].
Their values and symmetry are consistent with the LL ($c=1$ compact-boson) universality class~\cite{voit1995one}, whose vertex primary operators have scaling dimensions~\cite{nomura1997bkt}
\begin{equation}\label{delta}
  \Delta_{nm} = \frac{n^2}{2K}+\frac{m^2 K}{2}\qquad (n,m\in \mathbb{Z})
,\end{equation}
where $K\in\mathbb{C}$ is the Luttinger parameter.
We identify $n\bmod 3$ with the global $\mathbb{Z}_3$ charge, and $m$ with the winding number.
Under the convention $\operatorname{Re}(\beta)>0$, the cyan and purple regions satisfy $J>0$ and $J<0$, respectively.
We therefore refer to them as the ``FM-LL'' and ``AFM-LL'' phases.
These two regions are mapped to each other under the KW duality, and their Luttinger parameters satisfy $K(z)K(z^\star)=9$, as derived in \hyperref[endmatter]{End Matter}.
In the sandwiched pink region, Loop-TNR fails due to an extremely dense singular-value spectrum.
We resort to ED to compute finite-width ($L=12$) spectra at representative $z$ points [Fig.~\ref{fig2}(c)].
In the gapped phases, the degeneracy of $\lambda_0$ is consistent with the Loop-TNR spectra.
In the LL phases, near-dominant eigenvalues scatter sparsely, corresponding to scaling operators.
Remarkably, the sandwiched phase exhibits a ``snowflake'' spectrum distinct from both behaviors.
At $z=-0.7-0.3\mathrm{i}$, there are $L$ dominant eigenvalues with evenly spaced arguments, each accompanied by a dense cluster of near-dominant eigenvalues.
The spectrum at $z_\text{JRS}$ is more exotic, with all eigenvalues lying on the dominant circle.
This exact equimodularity has been noticed by JRS~\cite{jacobsen2006complex}, and is rigorously proved here for arbitrary $L$ and any boundary condition (see \hyperref[endmatter]{End Matter}).
Consequently, the system at $z_\text{JRS}$ is NCG, since all $D_\alpha=0$ is incompatible with the essential descendant structure in any CFT.
In the following, we will demonstrate numerically that the sandwiched phase is also NCG.
\begin{figure}[b!]
  \centering
  \includegraphics[width=\linewidth]{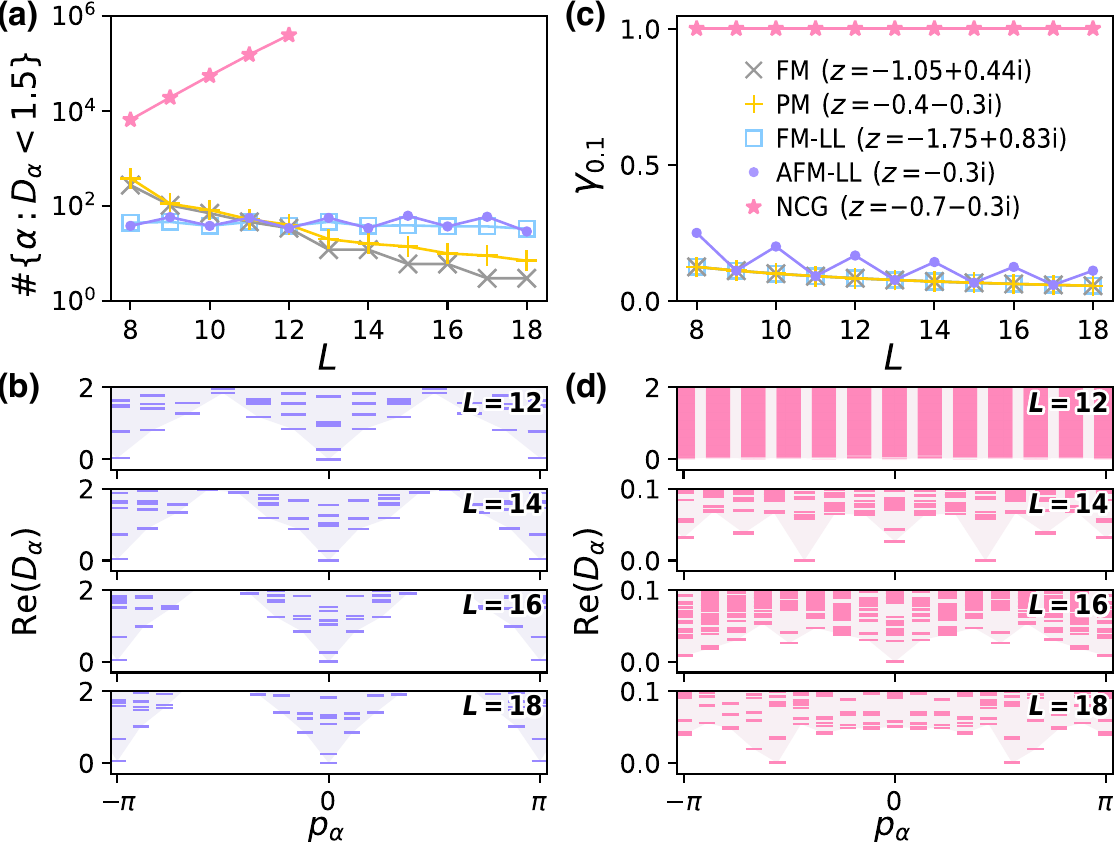}
  \caption{(a) The number of transfer-matrix eigenvalues with $D_\alpha < 1.5$. At the representative NCG point $z=-0.7-0.3\mathrm{i}$, only near-dominant eigenvalues are numerically accessible for $L>12$, so the corresponding counts are not shown. (b) Transfer-matrix levels evaluated at $z=-0.3\mathrm{i}$. Low-lying $D_\alpha$ stabilize to their corresponding $\operatorname{Re}(\Delta_\alpha)$, and the momentum distribution concentrates to $P=0$ and $\pi$. (c) The proportion of momentum sectors in which at least one eigenvalue satisfies $D_\alpha < 0.1$. (d) Transfer-matrix levels evaluated at $z=-0.7-0.3\mathrm{i}$. Panels for $L\ge 14$ are vertically zoomed in for clarity. No size-independent level structure emerges, and the momentum distribution does not concentrate to one or more centers.}
  \label{fig3}
\end{figure}
\textcolor{blue}{\textit{Non-Conformal Gaplessness}.}---We study the finite-size behavior of $\{D_\alpha\}$ at $z=-0.7-0.3\mathrm{i}$.
Figure~\ref{fig3}(a) shows that for the LL phases, the total number of eigenvalues satisfying $D_\alpha < 1.5$ has an upper bound, which is the expected behavior of compact CFTs.
In stark contrast, this number increases exponentially with $L$ at $z=-0.7-0.3\mathrm{i}$.
Although non-compact CFTs can have continuous scaling-dimension spectra and densifying $\{D_\alpha\}$, e.g., $D_\alpha \sim (\ln L)^{-2}$~\cite{ikhlef2008staggered}, it is still difficult to explain the exponential densification of $\{D_\alpha\}$.
\begin{figure}[b!]
  \centering
  \includegraphics[width=\linewidth]{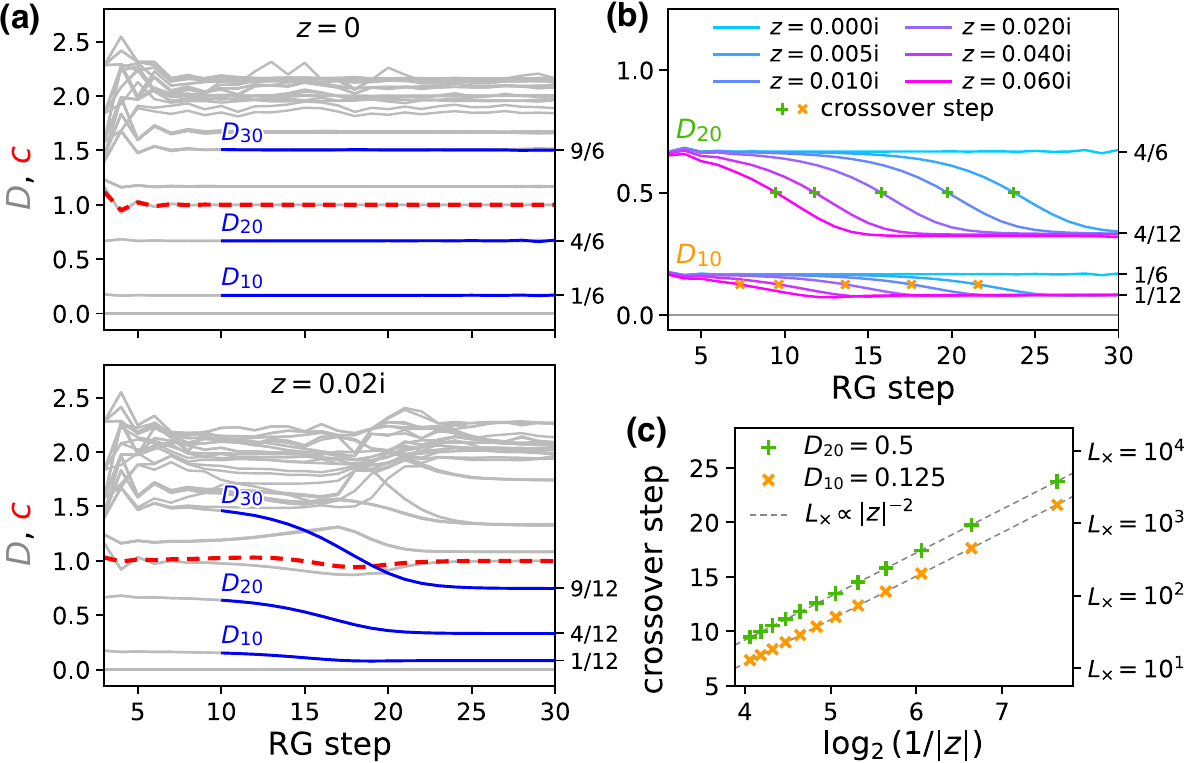}
  \caption{(a) Starting from $z=0$ and $0.02\mathrm{i}$, the RG flow terminates at LL fixed points with $K=3$ and $K\approx 6$, respectively. The transfer-matrix levels $D_{nm}$ corresponding to $\Delta_{nm}$ are highlighted in blue, which undergo a crossover at $z=0.02\mathrm{i}$. (b) As $z$ becomes closer to $0$, the crossover occurs at larger RG steps. (c) The RG step at which $D_{10}$ and $D_{20}$ intersect $0.125$ and $0.5$, respectively. The crossover lengths extracted from both gaps scale as $L_\times\sim \lvert z\rvert^{-2}$.}
  \label{fig4}
\end{figure}
We further resolve the transfer-matrix spectrum into $L$ momentum sectors.
We say that a transfer-matrix eigenvector has momentum $p$ if its eigenvalue under circumferential translation is $\mathrm{e}^{\mathrm{i}p}$, which satisfies $\mathrm{e}^{\mathrm{i}pL}=1$.
If the system is described by a CFT in the continuum limit, the distribution of $\{(D_\alpha,p_\alpha)\}$ is anchored by the conformal spectrum $\{(\Delta_\alpha, s_\alpha)\}$ as $p_\alpha = P + 2 \pi s_\alpha / L + o(L^{-1})$, where $s_\alpha$ is the conformal spin and $P$ is the momentum offset~\cite{cardy2010conformal}.
For instance, the AFM-LL phase has $P=0$ and $\pi$, with $\{(D_\alpha,p_\alpha)\}$ concentrating to them [Fig.~\ref{fig3}(b)].
Let $\gamma_{0.1}$ denote the proportion of momentum sectors in which at least one eigenvalue satisfies $D_\alpha<0.1$.
As shown in Fig.~\ref{fig3}(c), we have $\gamma_{0.1}\to 0$ in the LL phases, while $\gamma_{0.1}=1$ holds for every accessible $L$ at $z=-0.7-0.3\mathrm{i}$.
Figure~\ref{fig3}(d) further shows that this behavior is likely to persist as $L$ increases.
A CFT interpretation would therefore require an infinite number of primary operators satisfying $\operatorname{Re}(\Delta_\alpha-\Delta_0)<0.1$, and either their $\{s_\alpha\}$ is unbounded or their $\{P\}$ is dense throughout the Brillouin zone.
To our knowledge, no concrete CFT has such spectral features.
Together with the absence of a clear Casimir scaling of $\lambda_0$~\footref{sm}, these results provide strong evidence for an NCG phase.
\textcolor{blue}{\textit{Discontinuity at the AFM Point}.}---Having identified the AFM-LL phase, we now examine how it meets the AFM critical point.
Previous studies have shown that $z_\text{AFM}=0$ also belongs to the LL universality class with $K(0)=3$~\cite{salas1998three}, and that two scaling operators are excited under the thermal perturbation~\cite{cardy2001unusual}: one is a relevant vortex operator with RG eigenvalue $y_t = 1 / 2$, and the other is strictly marginal.
At positive temperature, the relevant operator prevents the RG flow from staying on the fixed line, so that $z_\text{AFM}$ does not extend to an LL phase.
Surprisingly, we find that under an imaginary perturbation, the RG flow initially leaves the LL fixed line but then returns to it at a \textit{distinct} fixed point.
At $z=0.02\mathrm{i}$, the Loop-TNR spectra remain close to $K=3$ at first, and then cross over to $K\approx6$ [Fig.~\ref{fig4}(a)].
By interpolating $D_\alpha$ between RG steps, we extract the crossover step and corresponding lattice size $L_\times$ at different $z$ points within the AFM-LL phase [Fig.~\ref{fig4}(b)].
They scale as $L_\times \sim \lvert z\rvert^{-2}$, consistent with $y_t = 1 / 2$ and implying that the crossover will eventually occur for arbitrarily small $\operatorname{Im}(z)>0$.
This establishes a twofold discontinuity $K(\mathrm{i}0^+) = 2 K(0) = 6$, reflecting the noncommutativity of the thermodynamic limit and $z\to \mathrm{i}0^+$.
Similar phenomena occur at the KW dual point $z_\text{AFM}^\star=-2$ where $K(-2+\mathrm{i}0^+)=\frac{1}{2}K(-2)=1.5$.
This jump can be understood from two known results.
First, the thermal perturbation at $z_\text{AFM}$ is described by a cosine potential added to the free-boson action, yielding the sine-Gordon model~\cite{delfino2001sine}.
Second, for the sine-Gordon model, when the coefficient of the cosine potential is imaginary, the massless RG flow takes the Gaussian coupling from $g_\text{UV}$ to $g_\text{IR}=2-g_\text{UV}$~\cite{fendley1993massless}.
Using our convention $K=4 / g$, this corresponds to a flow from $K_\text{UV}=3$ to $K_\text{IR}=6$.
In the $K\approx 6$ regime, the aforementioned vortex operator becomes irrelevant and the LL phase is stabilized.
\textcolor{blue}{\textit{Implications and Outlook}.}---By resolving the global Fisher zero distribution, our results unify and revise long-standing conjectures and establish a phase diagram featuring both conformal and non-conformal gapless phases.
The two LL phases provide a concrete realization of the KW duality relating extended critical phases, beyond its familiar role in relating ordered and disordered phases.
Recent studies have also shown that complex-temperature phases may represent metastable phases at real temperature, which are encoded in the partition function but absent from the real-temperature equilibrium phase diagram~\cite{dong2026tracking}.
This motivates asking whether the AFM-LL phase has a metastable continuation to real temperature, and whether it can be stabilized by introducing suitable real-valued couplings.
The rich phase structure also offers targets for quantum experiments.
Quantum circuit protocols for measuring complex-temperature partition functions explicitly apply to two-dimensional Ising and Potts models~\cite{krishnan2019measuring}.
It has been demostrated that dense Fisher zeros of the anisotropic Ising model correspond to a spatially or temporally modulated phase~\cite{basu2022fisher}.
Given recent measurements of Fisher zeros on quantum hardware~\cite{laba2025quantum,kuzmak2025detecting}, we expect this route to enable experimental realizations of the complex-temperature Potts model and its gapless phases.
\begin{acknowledgments}
\textcolor{blue}{\textit{Acknowledgements}.}---We thank Yi-Hua Dong, Ling Liu, Jia-Cheng Gao, Qing-Yang Zheng, and Wuji Han for helpful discussions.
We are supported by the Science Challenge Project under Grant No.~TZ02052013, the National Science Foundation of China under Grants No.~123B2048, 12522410, 12550005, 12234001, 12474215, 62321004, and the National Key Laboratory of Computational Physics under Grant No.~SYSM-2026-WDZC-04.
The computational resources were provided by the supercomputer center at Peking University, China.
\end{acknowledgments}

\begin{widetext}
	\section{End Matter} \label{endmatter}
\end{widetext}
\textcolor{blue}{\textit{Appendix A: Transfer Matrix}.}---Throughout the appendices, addition involving spins $\sigma_j$ should be understood modulo 3.
Denote the configuration of a lattice row by $\boldsymbol{\sigma}\equiv (\sigma_1,\cdots,\sigma_L)$ and the transfer matrix by $\mathcal{T}^{(\tau)}$, where $\tau\in\{0,1,-1\}$ defines the \textit{twisted} boundary condition $\sigma_{L+1} = \sigma_1+\tau$.
The transfer-matrix element is determined by the energy change of adding a new row $\boldsymbol{\sigma}$ adjacent to $\boldsymbol{\sigma}'$, as
\begin{equation}
	\langle  \boldsymbol{\sigma} | \mathcal{T}^{(\tau)} | \boldsymbol{\sigma}' \rangle \equiv \mathrm{e}^{-\beta [- JU^{(\tau)}(\boldsymbol{\sigma}) - JV(\boldsymbol{\sigma},\boldsymbol{\sigma}')]}\label{tm}
,\end{equation}
where
\begin{align}
	U^{(\tau)}(\boldsymbol{\sigma})\equiv \sum_{j=1}^{L} \delta_{\sigma_j\sigma_{j+1}},\quad V(\boldsymbol{\sigma},\boldsymbol{\sigma}')\equiv \sum_{j=1}^{L} \delta_{\sigma_j\sigma_j'}
\end{align}
are the numbers of satisfied bonds within a single row and between two adjacent rows, respectively.
The spectra shown in Figs.~\ref{fig2}--\ref{fig4} are computed using the periodic boundary condition ($\tau=0$).
Note that $\mathcal{T}^{(0)}$ commutes with the circumferential translation $\mathcal{J}$ and the cyclic spin-shift $\mathcal{G}$, which are defined as
\begin{gather}
	\mathcal{J} \left| \boldsymbol{\sigma} \right> = \left| (\sigma_L, \sigma_1, \cdots, \sigma_{L-1}) \right>,
	\\\mathcal{G} \left| \boldsymbol{\sigma} \right> = \left| \sigma_1+1, \sigma_2+1, \cdots, \sigma_L+1 \right>
.\end{gather}
They satisfy $\mathcal{J}^L=\mathcal{G}^3=\mathcal{I}$, where $\mathcal{I}$ is the identity matrix.
Therefore, the eigenvalues of $\mathcal{T}^{(0)}$ can be grouped into $L\times 3$ sectors labeled by $(p,Q)$, where $p$ is the momentum and $Q\in\{0,1,-1\}$ is the global $\mathbb{Z}_3$ charge.
As for $\mathcal{T}^{(\pm 1)}$, they also commute with $\mathcal{G}$ so that $Q$ is still well-defined.
Figure~S1 shows the charge-resolved spectra of $\mathcal{T}^{(\tau)}$ computed via Loop-TNR.
The low-lying levels are consistent with
\begin{equation}
	\Delta_{nm}^{(\tau)} = \frac{n^2}{2K} + \frac{K}{2}\left(m+\frac{\tau}{3}\right)^2\pod{n,m\in\mathbb{Z}}\label{deltatau}
,\end{equation}
identifying the quantum number $m$ as the integer winding.
They also satisfy $Q=n\bmod 3$, implying that the spin-shift $\mathcal{G}$ acts as a shift on the compact boson by one third of its compactification period.
\textcolor{blue}{\textit{Appendix B: The KW Duality}.}---Given two plane graphs $G$ and $G^\star$ that are dual graphs of each other, the KW duality refers to the exact relation~\cite{wu1982potts}
\begin{equation}
	Z_G(z) = 3^{N(G)-b(G)-1} (z-1)^{b(G)} Z_{G^\star}(z^\star)
,\end{equation}
where $Z_G$ is the partition function of the 3-state Potts model defined on $G$, $N(G)$ is the number of vertices, and $b(G)$ is the number of edges.
The infinite-size square lattice is a self-dual graph, so that
\begin{align}
	f(z)&=-\lim_{N\to \infty} \frac{1}{N}\ln Z(z) 
	  \\&=-\lim_{N\to \infty} \frac{1}{N}\ln \left[3^{-N-1}(z-1)^{2N} Z(z^\star)\right]
	  \\&= f(z^\star) - 2\ln ( z-1 ) + \ln 3\label{selfdual}
.\end{align}
The KW duality applies not only to $Z(z)$, but also to transfer matrices.
Since $\mathcal{T}^{(\tau)}$ commutes with $\mathcal{G}$, it can be decomposed into three blocks as $\mathcal{T}^{(\tau)}=\bigoplus_Q \mathcal{T}^{(\tau,Q)}$.
Based on the isomorphism proposed by Han and Honecker~\cite{han1994low}, we can prove that for any $z\neq 1$~\footref{sm},
\begin{equation}
\mathcal{T}^{(\tau,Q)}(z^\star) \simeq 3^{L}(z-1)^{-2L}\mathcal{T}^{(Q,\tau)}(z)
,\end{equation}
where $\simeq$ denotes the matrix similarity.
Therefore, the respective dominant eigenvalues of $\mathcal{T}^{(0,0)}(z)$, $\mathcal{T}^{(0,0)}(z^\star)$, $\mathcal{T}^{(1,0)}(z)$, and $\mathcal{T}^{(0,1)}(z^\star)$ satisfy
\begin{align}\label{eig0}
	\lambda_0^{(0,0)}(z^\star) &= 3^L (z-1)^{-2L} \lambda_0^{(0,0)}(z),\\
	\lambda_0^{(0,1)}(z^\star) &= 3^L (z-1)^{-2L} \lambda_0^{(1,0)}(z)\label{eig1}
.\end{align}
Combining Eqs.~(\ref{deltatau})(\ref{eig0})(\ref{eig1}) and taking $L\to\infty$ yields
\begin{equation}
	\Delta_{10}^{(0)}(z^\star) = \Delta_{00}^{(1)}(z) = \frac{1}{2 K(z^\star)} = \frac{K(z)}{18}
,\end{equation}
and hence $K(z)K(z^\star)=9$.
\textcolor{blue}{\textit{Appendix C: The JRS Point}.}---The transfer matrix defined in Eq.~(\ref{tm}) can be written as $\mathcal{T}^{(\tau)}=\mathcal{U}^{(\tau)}\mathcal{V}$, where
\begin{equation}
	\mathcal{U}^{(\tau)} \left| \boldsymbol{\sigma} \right> = z^{U^{(\tau)}(\boldsymbol{\sigma})} \left| \boldsymbol{\sigma} \right>
\end{equation}
is diagonal, and
\begin{equation}
	\mathcal{V} = \begin{bmatrix}
	  z&1&1\\1&z&1\\1&1&z
  \end{bmatrix}^{\otimes L}
.\end{equation}
Note that $z_\text{JRS}=(-1\pm \sqrt{3}\mathrm{i}) / 2$ is a primitive cube root of unity, which we denote by $\omega$ for simplicity.
At $z=\omega$, we have $\mathcal{U}^{(\tau)}{}^\dagger \mathcal{U}^{(\tau)} = \mathcal{I}$, since
\begin{equation}
	\overline{z}^{U^{(\tau)}(\boldsymbol{\sigma})}z^{U^{(\tau)}(\boldsymbol{\sigma})} = (\overline{\omega}\omega)^{U^{(\tau)}(\boldsymbol{\sigma})} = 1
,\end{equation}
and $\mathcal{V}^\dagger \mathcal{V} = 3^L \mathcal{I}$, since
\begin{equation}
  \begin{bmatrix}
	  \omega&1&1\\1&\omega&1\\1&1&\omega
  \end{bmatrix}^\dagger \begin{bmatrix}
	  \omega&1&1\\1&\omega&1\\1&1&\omega
  \end{bmatrix} = \begin{bmatrix}
	  3&0&0\\0&3&0\\0&0&3
  \end{bmatrix}
.\end{equation}
Consequently, $3^{-L / 2}\mathcal{T}^{(\tau)}$ is unitary, so that its all eigenvalues lie on the unit circle.
So far, we have proved that at $z=z_\text{JRS}$, all transfer-matrix eigenvalues satisfy $\lvert \lambda_\alpha \rvert = 3^{L / 2}$ regardless of $L$ and $\tau$.
In fact, this property holds for \textit{any} boundary conditions (e.g., the open boundary condition), since they only affect $U^{(\tau)}(\boldsymbol{\sigma})$ which remains real.

\end{document}


\title{Supplemental Material for ``Three New Phases in the Two-Dimensional Potts Model at Complex Temperature''}
\maketitle
\tableofcontents
\section{Extracting complex scaling dimensions from Loop-TNR results}
\begin{figure}[b!]
  \centering
  \includegraphics[width=\linewidth]{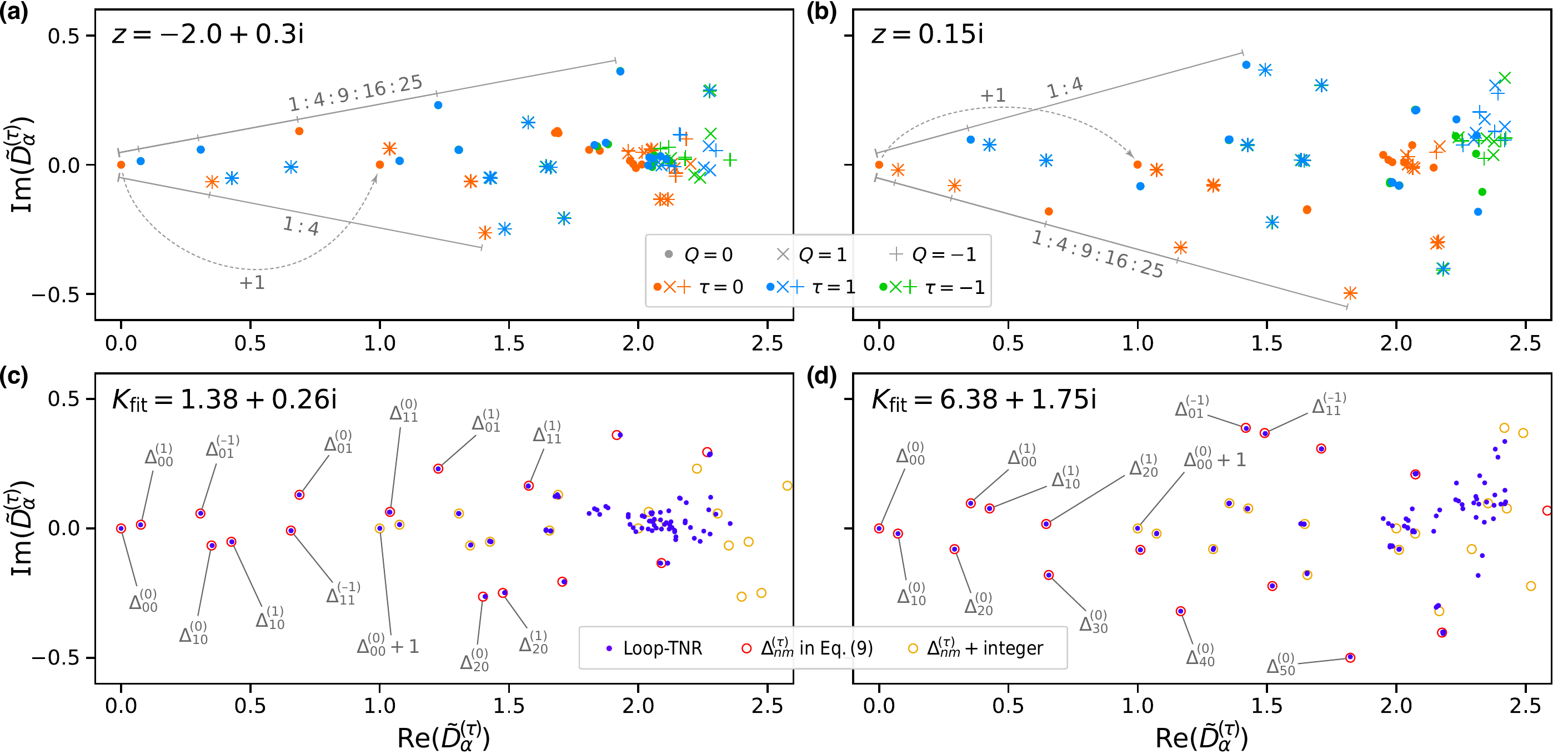}
  \caption{(a)(b) Complex transfer-matrix levels $\tilde D_\alpha^{(\tau)}$ extracted from Loop-TNR results. They exhibit integer ratios (solid gray lines), descendant structures (dashed gray lines), and parallelogram patterns. Theoretically, the spectra of $\{D_\alpha^{(1)}\}$ and $\{D_\alpha^{(-1)}\}$ are identical by virtue of the $\sigma_j\mapsto -\sigma_j$ symmetry, and so are the $Q=\pm 1$ sectors. However, they differ when Loop-TNR fails to capture higher scaling dimensions due to the finiteness of $\chi$. For $\tau=0$, there is a family of $D_\alpha^{(0)}\propto n^2$ and $Q=n\bmod 3$. For the $Q=0$ sector, there is also a family of $D_\alpha^{(\tau)}\propto m^2$ and $\tau=m\bmod 3$. (c)(d) Complex scaling dimensions $\Delta_{nm}^{(\tau)}$ (red circles) of the LL predicted by Eq.~(9) and their descendants (yellow circles). They are consistent with the low-lying $D_\alpha^{(\tau)}$ (purple dots) obtained via Loop-TNR.}
  \label{figs1}
\end{figure}
Using Loop-TNR results, we construct the transfer matrix via the configuration depicted in Fig.~3(g) of Ref.~[\citenum{yang2017loop}].
When the RG step $\ell$ is even, this transfer matrix equals the $L$-th power of the single-row transfer matrix $\mathcal{T}^{(\tau)}$, where $L=2^{\ell / 2}$.
We use its eigenvalues $\{(\lambda_\alpha^{(\tau)})^L\}$ to compute the complex-valued level
\begin{equation}
	\tilde{D}_\alpha^{(\tau)} \equiv \frac{1}{2\pi} \ln \frac{(\lambda_0^{(0)})^L}{\mathrm{e}^{-\mathrm{i}PL}(\lambda_\alpha^{(\tau)})^L}
,\end{equation}
where the momentum offset $P$ can be determined from momentum-resolved ED spectra.
In the two LL phases, we have $P\in\{0,\pi\}$, so that $\mathrm{e}^{-\mathrm{i}P L}=1$ holds for all $\ell=2,4,6,\cdots$.
After choosing the right branch of the logarithm, we expect $\tilde{D}_\alpha^{(\tau)} \to \Delta_\alpha^{(\tau)} - \Delta_0^{(0)}$ as $L\to\infty$.
However, since the bond dimension $\chi$ is finite, $\tilde{D}_\alpha^{(\tau)}$ must be computed at a moderate RG step $\ell$, not as large as possible~\cite{chan2026extracting}.
Figure~\ref{figs1} shows the low-lying $\{\tilde{D}_\alpha^{(\tau)}\}$ computed at representative points $z=-2.0+0.3\mathrm{i}$ and $z=0.15\mathrm{i}$ with $\ell=20$, and the $\{\Delta_{nm}^{(\tau)}\}$ predicted by Eq.~(9) for comparison.
Their consistency strongly supports the LL description, and the quantum numbers $n$ and $m$ are related to $Q$ and $\tau$, respectively.
The Luttinger parameter $K_\text{fit}$ is determined using the twelve $\tilde{D}_\alpha^{(\tau)}$ with smallest real parts.
\begin{figure}[t!]
  \centering
  \includegraphics[width=\linewidth]{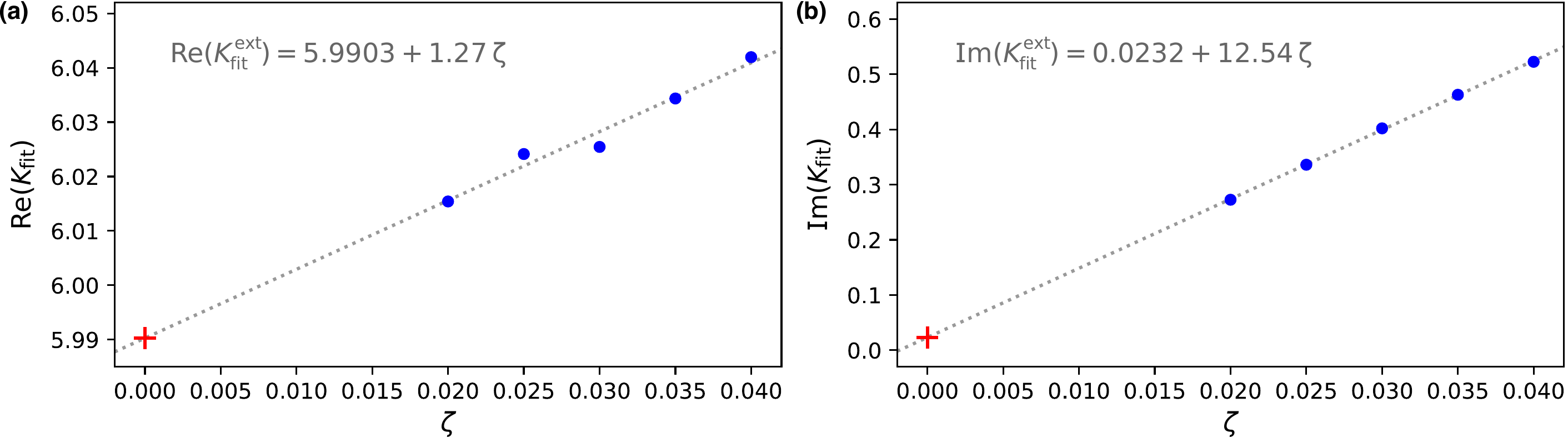}
  \caption{Luttinger parameters (blue dots) fitted from Loop-TNR results and the extrapolated value at $\zeta=0$ (red crosses).}
  \label{figs2}
\end{figure}
The limiting value of the Luttinger parameter $K(\mathrm{i}0^+)$ is extrapolated using $K_\text{fit}(\mathrm{i}\zeta)$ from $\zeta=0.020$ to $0.040$, fitted from low-lying $\{\tilde{D}_\alpha^{(0)}\}$ at $\ell=28$ [Fig.~\ref{figs2}].
The resulting $K_\text{fit}^\text{ext}(0)\approx 5.99+0.02\mathrm{i}$ is close to $2K(0)=6$.
We can also see that increasing $\zeta$ mainly contributes to $\operatorname{Im}(K)$ rather than $\operatorname{Re}(K)$.
\section{Identifying and suppressing finite-entanglement artifacts of BWTRG}
\begin{figure}[t!]
  \centering
  \includegraphics[width=\linewidth]{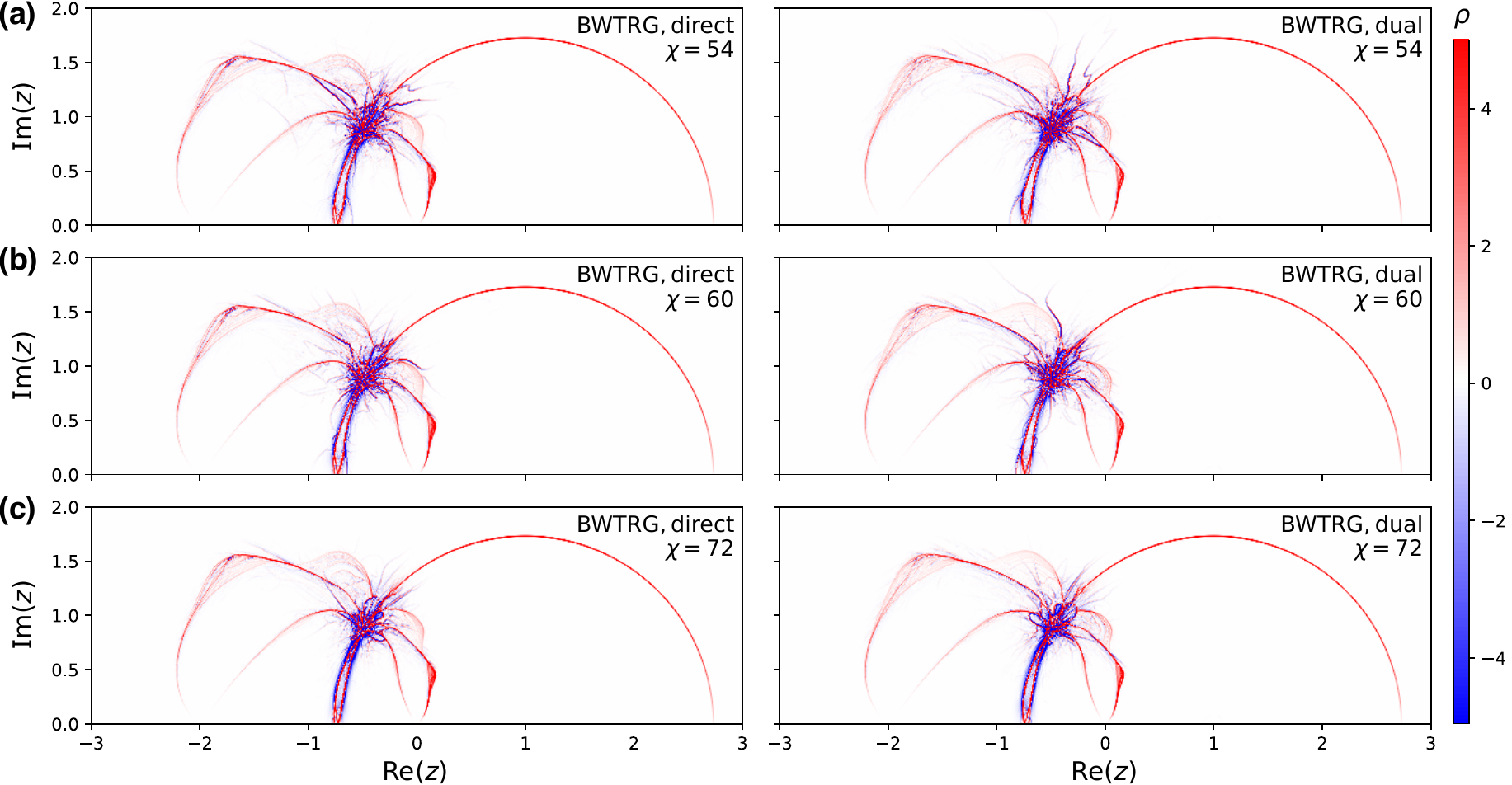}
  \caption{$\rho_\text{direct}$ and $\rho_\text{dual}$ computed by BWTRG at (a) $\chi=54$, (b) $\chi=60$, and (c) $\chi=72$. One can see that most finite-entanglement artifacts are not self-dual and appear as adjacent positive-negative pairs.}
  \label{figs3}
\end{figure}

\begin{figure}[b!]
  \centering
  \includegraphics[width=\linewidth]{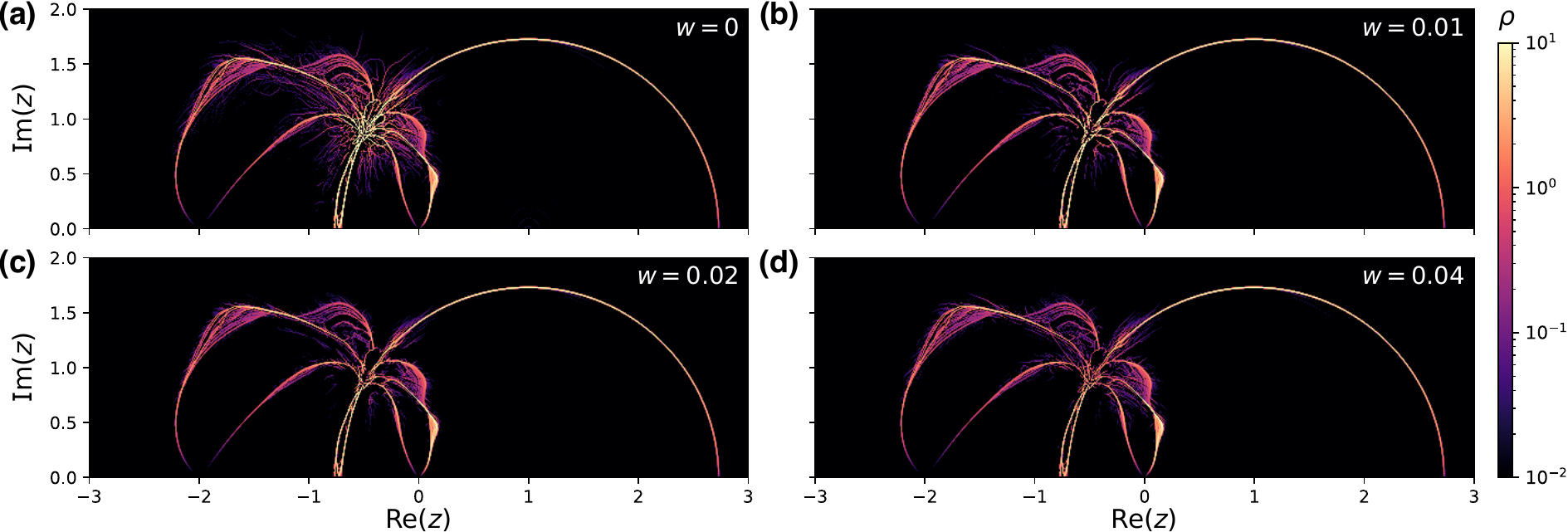}
  \caption{$\hat{\rho}_\text{direct}$ computed at $\chi=72$ and filtered using different $w$ values. Most ringing artifacts are suppressed when $w=0.02$.}
  \label{figs4}
\end{figure}

The zero density $\rho(z)$ in Fig.~1(c) is computed via the following precedure.
We define a square grid on the region $\{z\in\mathbb{C} : -3\le\operatorname{Re}(z)\le 3,\;0\le\operatorname{Im}(z)\le 2\}$ with grid spacing $h=0.01$.
For each $z$ point in the grid, we invoke BWTRG twice.
On the one hand, we construct the initial local tensor $A(z)$ of BWTRG, which directly yields the free energy $f_\text{direct}(z)$.
On the other hand, we construct the initial local tensor $A(z^\star)$, yielding $f_\text{direct}(z^\star)$ and
\begin{equation}
	f_\text{dual}(z)\equiv f_\text{direct}(z^\star)-2\ln (z-1) +\ln q
.\end{equation}
According to Eq.~(13), we have $f_\text{direct}=f_\text{dual}$ theoretically, but they are numerically different since $\chi$ is finite.
We use them to compute $\rho_\text{direct}$ and $\rho_\text{dual}$, respectively, via the discrete Laplacian
\begin{equation}
  \rho(z) = - \frac{1}{2\pi} \nabla^2 \operatorname{Re} f(z) \approx \frac{1}{2\pi} \operatorname{Re} \frac{4f(z)-f(z+h)-f(z-h)-f(z+\mathrm{i}h)-f(z-\mathrm{i}h)}{h^2}
.\end{equation}
Figure~\ref{figs3} shows the results with $\chi=54,60,72$.
We distinguish true Fisher zero structures from finite-entanglement artifacts by three criteria: (i) they satisfy $\rho\ge 0$, (ii) they are stable under increasing $\chi$, and (iii) they should appear in both $\rho_\text{direct}$ and $\rho_\text{dual}$.
Using these criteria, we find that finite-entanglement artifacts mainly appear around $z_\text{JRS}$ as \textit{ringing artifacts}, i.e., adjacent positive-negative pairs.
To supress the ringing artifacts, we apply a nonlinear filter to $\rho_\text{direct}$ and $\rho_\text{dual}$, respectively.
We define the credibility of each $z$ point in the grid as
\begin{equation}
	\kappa (z;w) \equiv \frac{\sum_{j,k} \mathrm{e}^{-(j^2+k^2)h^2 / 2w^2} \rho(z+jh+\mathrm{i}kh)}{\sum_{j,k} \mathrm{e}^{-(j^2+k^2)h^2 / 2w^2} \lvert\rho(z+jh+\mathrm{i}kh)\rvert}\label{cred}
,\end{equation}
where $w>0$ is the standard deviation of the Gaussian filter.
The denominator of Eq.~(\ref{cred}) quantifies the average $\lvert\rho\rvert$ near $z$, while the numerator reflects the \textit{net} zero density.
When $\rho\ge 0$, we have $\kappa(z)=1$.
Near the ringing artifacts, if $w$ is sufficiently large to cover the positive-negative pattern, the terms in the numerator of Eq.~(\ref{cred}) will cancel and we expect $\kappa(z) \approx 0$.
Therefore, we define
\begin{equation}
  \hat{\rho}(z;w) \equiv \begin{cases}
	  \kappa(z;w) \rho(z) & \text{if }\kappa(z;w)>0\text{ and }\rho(z)>0,\\
	  0&\text{otherwise}.
  \end{cases}
\end{equation}
As shown in Fig.~\ref{figs4}, using $w=0.02$ suppresses most ringing artifacts, and preserves the stable features in Fig.~\ref{figs3}.
In the main text, Fig.~1(c) actually shows $(\hat\rho_\text{direct}+\hat\rho_\text{dual})/2$ with $w=0.02$.
\section{Fisher zero density obtained with different methods}
\begin{figure}[b!]
  \centering
  \includegraphics[width=\linewidth]{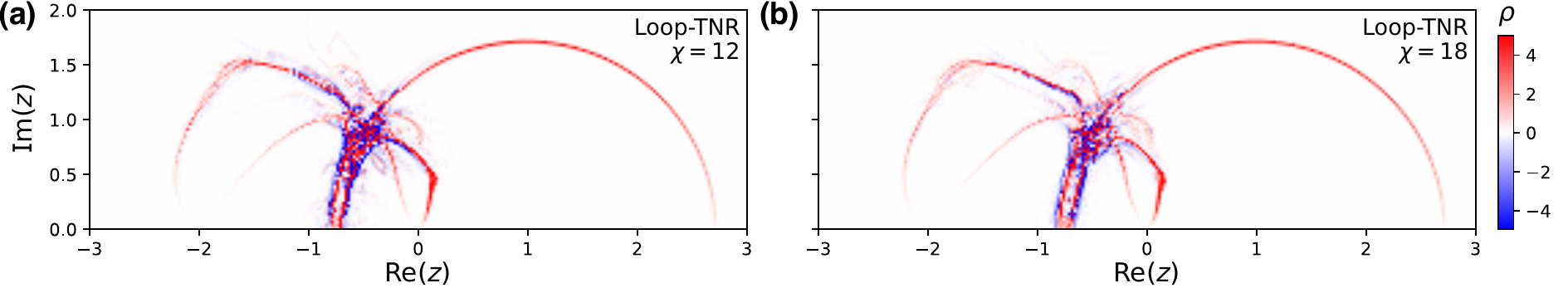}
  \caption{$\rho_\text{direct}$ computed by Loop-TNR at (a) $\chi=12$ and (b) $\chi=18$.}
  \label{figs5}
\end{figure}
\begin{figure}[b!]
  \centering
  \includegraphics[width=\linewidth]{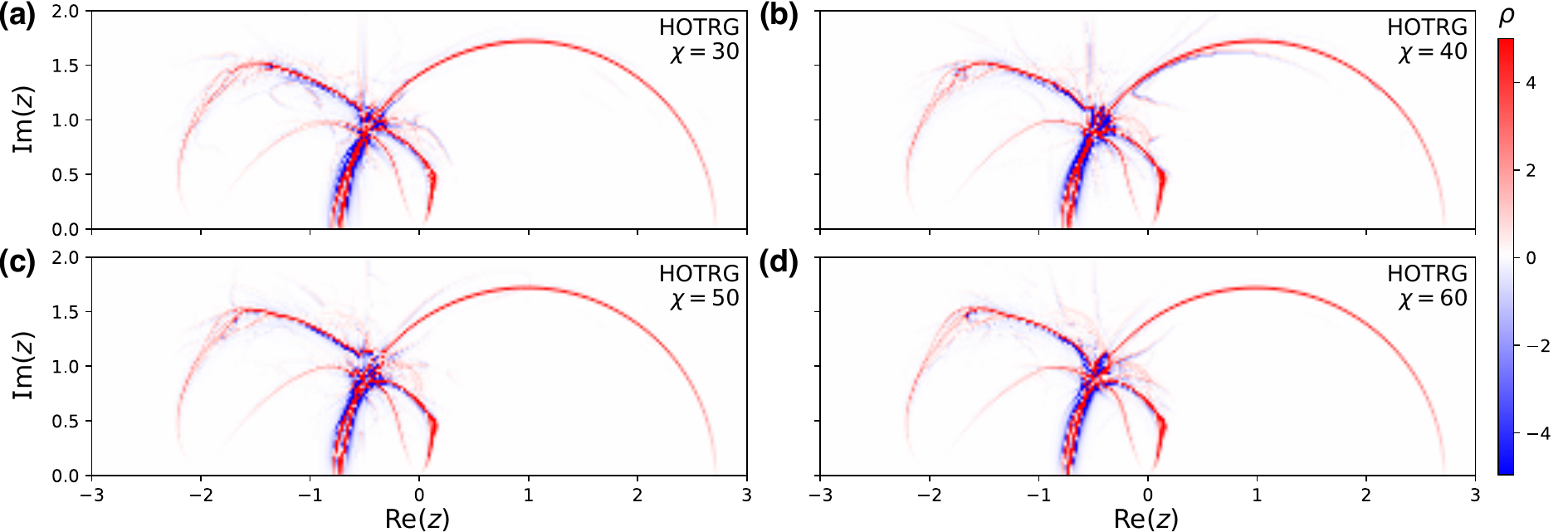}
  \caption{$\rho_\text{direct}$ computed by HOTRG at (a) $\chi=30$, (b) $\chi=40$, (c) $\chi=50$, and (d) $\chi=60$.}
  \label{figs6}
\end{figure}
\begin{figure}[b!]
  \centering
  \includegraphics[width=\linewidth]{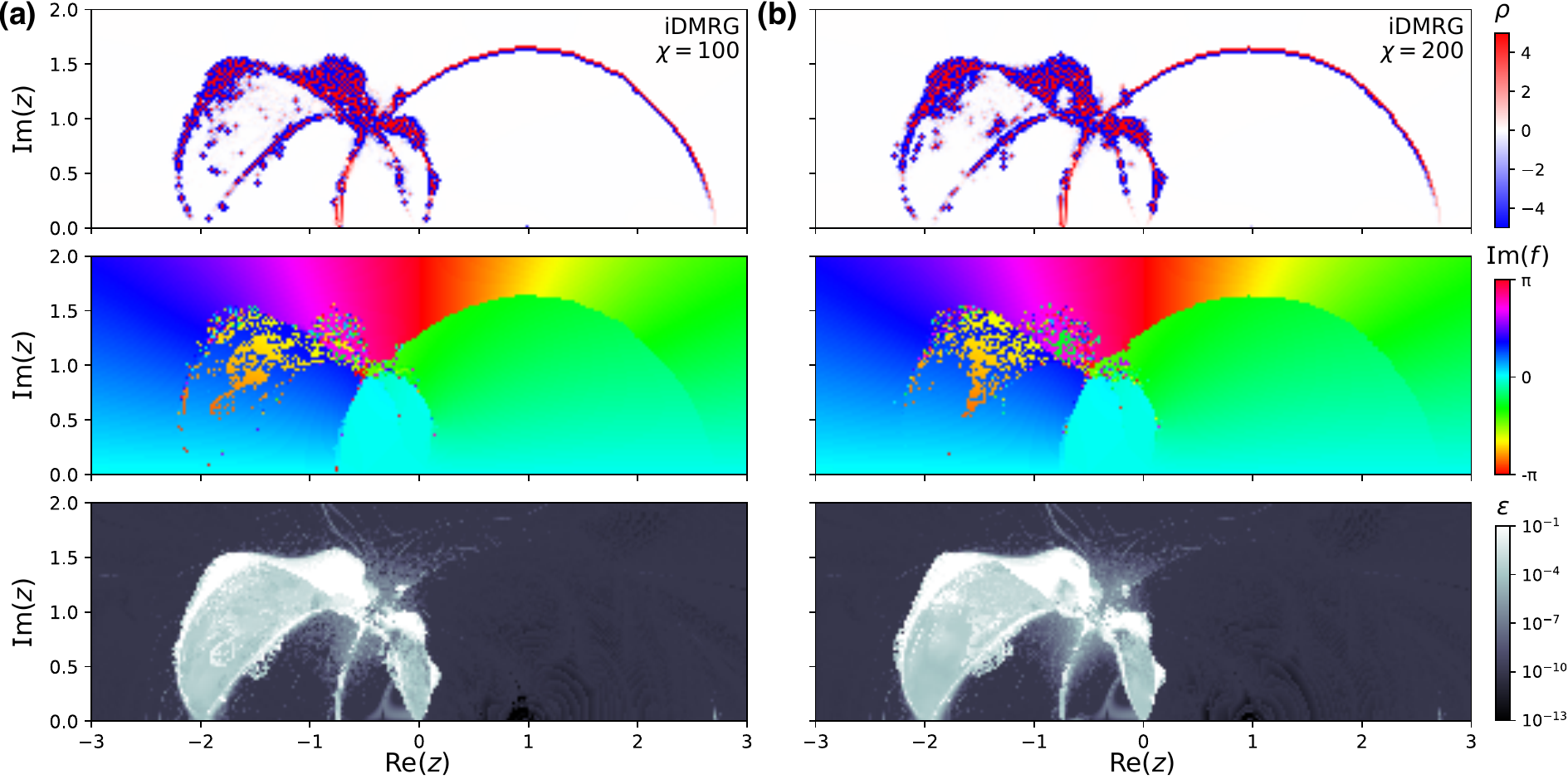}
  \caption{$\rho_\text{direct}$, $\operatorname{Im}f$, and the Galerkin error $\varepsilon$ computed by iDMRG with $2000$ iteration steps at (a) $\chi=100$ and (b) $\chi=200$. Since iDMRG relies on the random initial tensor, it may converge to wrong branches and yield discontinuous $f(z)$, especially in gapless phases and dense-zero areas. However, $\varepsilon(z)$ exhibits a rather stable ``phase diagram'' pattern.}
  \label{figs7}
\end{figure}
\begin{figure}[t!]
  \centering
  \includegraphics[width=\linewidth]{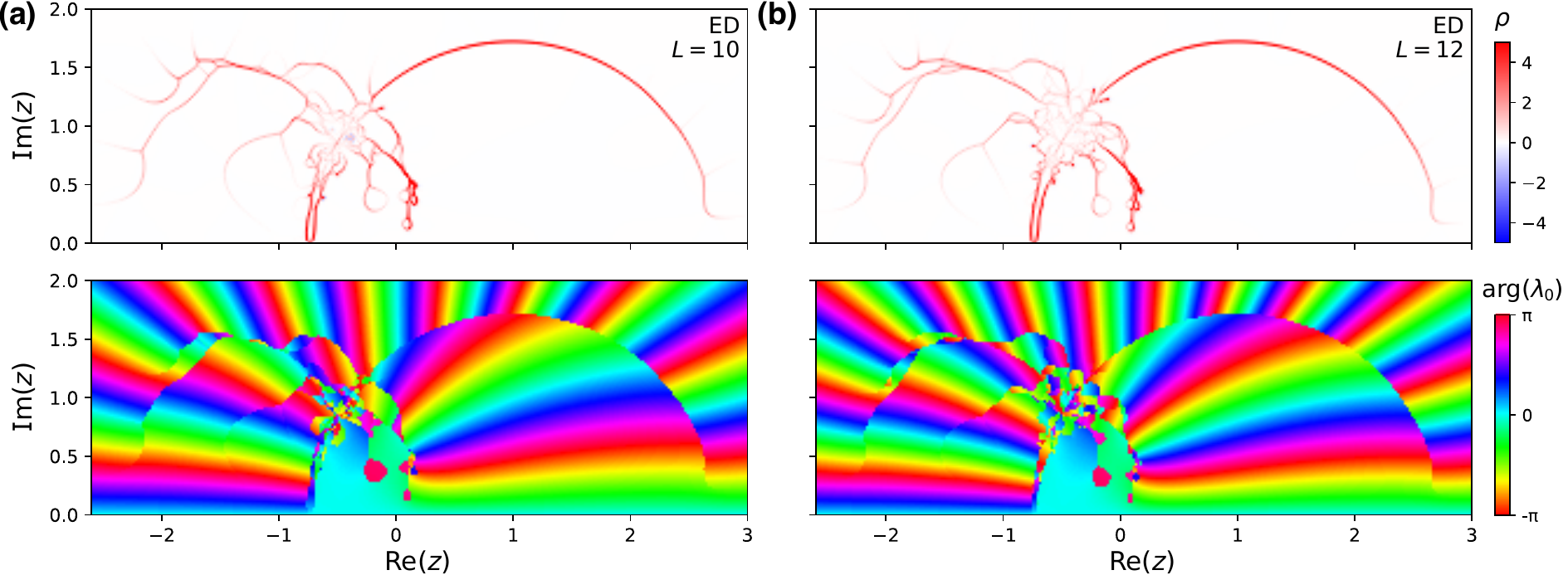}
  \caption{$\rho_\text{direct}$ computed by $-\frac{1}{L}\ln \lvert \lambda_0(z) \rvert \approx \operatorname{Re}f(z)$ at (a) $L=10$ and (b) $L=12$. The argument of the dominant eigenvalue are shown below, which approximates $-L\operatorname{Im}f(z)\bmod 2\pi$.}
  \label{figs8}
\end{figure}

We have shown $\rho(z)$ computed by BWTRG in Figs.~1(c), \ref{figs3}, and \ref{figs4}.
Now, we present $\rho(z)$ obtained from iDMRG, Loop-TNR, and higher-order tensor renormalization group (HOTRG)~\cite{xie2012coarse} with $z$-grid spacing $h=0.025$.
Similar to BWTRG, Loop-TNR and HOTRG are robust and directly produce $\operatorname{Re}f(z)$.
We also obtain $\operatorname{Im}f(z)$ via iDMRG which produces $\mathrm{e}^{- f(z)}$, but it is hard to converge in the gapless phases---the well-converged result in Fig.~1(d) is obtained using ${\sim}10^4$ iteration steps, smaller $\chi$, and multiple trial initial tensors.
Moreover, we use ED to compute $\lambda_0(z)$ with $z$-grid spacing $h=0.02$, and use $-\frac{1}{L}\ln \lvert\lambda_0(z)\rvert$ to approximate $\operatorname{Re}f(z)$ since~\cite{blote1986conformal}
\begin{equation}
  -\ln \lvert\lambda_0(z)\rvert = \begin{cases}
	  L \operatorname{Re}f(z) + O(\mathrm{e}^{- L / \xi})&\text{if the system is gapped},\\
	  L \operatorname{Re}f(z) - \dfrac{\pi}{6L}\operatorname{Re}(c) + o(L^{-1})&\text{if the system is conformal gapless},
  \end{cases}\label{casimir}
\end{equation}
where $c$ is the \textit{effective} central charge.
Figures~\ref{figs5}--\ref{figs8} show $\rho(z)$ obtained by these four methods, respectively.
They are consistent with the phase diagram in Fig.~2(a).
The ringing artifacts seems to be a common trait of tensor network renormalization methods (BWTRG, Loop-TNR, and HOTRG), while the finite-size artifact in Fig.~\ref{figs8}(a)(b) are mostly nonnegative.
We also notice that the Galerkin error $\varepsilon(z)$ of iDMRG clearly distinguishes the gapped phase, the gapless phase, and the dense-zero area.
\section{Finite-size scaling of the dominant transfer-matrix eigenvalue}
\begin{figure}[b!]
  \centering
  \includegraphics[width=\linewidth]{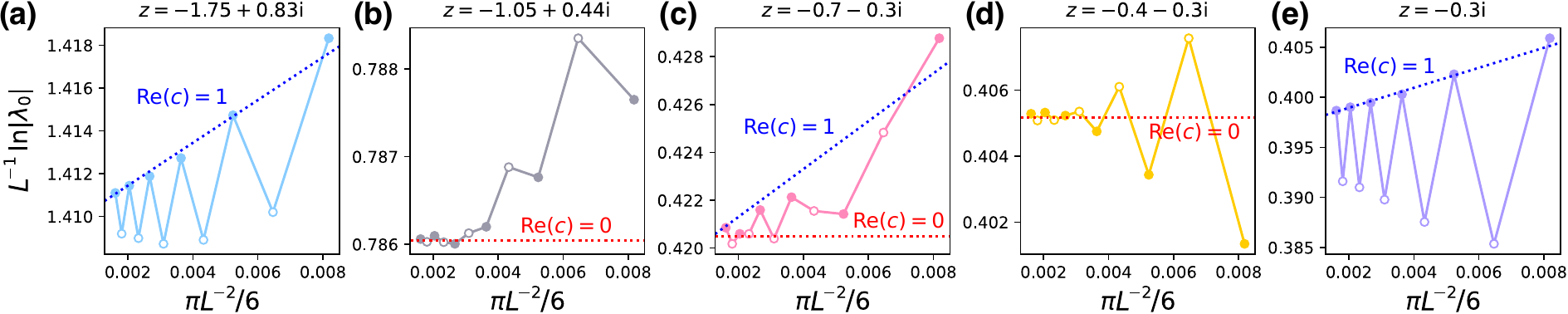}
  \caption{Finite-size scaling of $\lvert\lambda_0\rvert$ evaluated at representative $z$ points. Solid and hollow symbols represent even and odd $L$, respectively. Dashed lines with slopes $\operatorname{Re}(c)=1$ and $0$ are only used for reference. Panels (a) and (e) shows the conformal behavior of the LL phases, while (b) and (d) shows the gapped behavior of the FM and PM phases. In Panel (c), we cannot identify a conformal subsequence with a steady slope. Even if such slope exists, it seems to have $\operatorname{Re}(c)$ strictly less than $1$.}
  \label{figs9}
\end{figure}
If the NCG phase has a CFT description, the dominant eigenvalue $\lambda_0$ should have the Casimir scaling in Eq.~(\ref{casimir}).
Figure~\ref{figs9} shows $\frac{1}{L}\ln \lvert \lambda_0\rvert $ for $L\le 18$ at representative $z$ points.
For the LL phases, the untwisted sector ($\tau=0$) is realized at even $L$ with $c=1$.
For the gapped phases, the finite-size effect vanishes rapidly, exhibiting the $c=0$ behavior.
The NCG phase is different from both cases.
If we manage to fit a value of $\operatorname{Re}(c)$, it will lie substantially below $1$, which usually suggests a compact CFT.
However, the compact-CFT description is challenged in the main text since they cannot account for the exponentially densifying transfer-matrix levels with $D_\alpha<1.5$.
\section{Proof for the KW duality of transfer matrices}
In this section, we will prove the statement in Eq.~(14) that $\mathcal{T}^{(\tau,Q)}(z^\star )$ is similar to $3^L(z-1)^{-2L}\mathcal{T}^{(Q,\tau)}(z)$ for any $L\in \mathbb{Z}_+$ and $z\neq 1$, basically rephrasing the work of Han and Honecker~\cite{han1994low} in the language of classical 3-state Potts model.
Let us define the local spin-shift matrix $\mathcal{X}_j$ and the spin-difference matrix $\mathcal{B}^{(\tau)}_j$ as
\begin{gather}
  \mathcal{X}_j \left| \boldsymbol{\sigma} \right> \equiv \left| \sigma_1, \cdots, \sigma_{j-1}, \sigma_j + 1, \sigma_{j+1}, \cdots, \sigma_L \right>,\\
  \mathcal{B}_j^{(\tau)} \left| \boldsymbol{\sigma} \right> \equiv \omega^{\sigma_{j+1} - \sigma_j} \left| \boldsymbol{\sigma} \right>
,\end{gather}
where $\omega\equiv \mathrm{e}^{2\pi\mathrm{i} / 3}$, $j\in\{1,\cdots,L\}$, and $\mathcal{B}_j^{(\tau)}$ depends on $\tau$ only when $j=L$.
Note that they satisfy
\begin{gather}
	\left(\mathcal{I}+\mathcal{B}_j^{(\tau)}+(\mathcal{B}_j^{(\tau)})^2\right)\left| \boldsymbol{\sigma} \right> = 3 \delta_{\sigma_j\sigma_{j+1}}\left| \boldsymbol{\sigma} \right>,\\
	\mathcal{I}+\mathcal{X}_j+\mathcal{X}_j^2 = \mathcal{I}_1^{\otimes (j-1)} \otimes \begin{bmatrix}
		1&1&1\\1&1&1\\1&1&1
	\end{bmatrix} \otimes \mathcal{I}_1^{\otimes (L-j)}
,\end{gather}
where $\mathcal{I}_1$ denotes the $3\times 3$ identity matrix.
The matrices $\mathcal{U}^{(\tau)}(z)$ and $\mathcal{V}(z)$ defined in Eqs.~(18) and (19) can be rewritten as
\begin{gather}
	\mathcal{U}^{(\tau)}(z)
	=\prod_{j=1}^L\left[\mathcal{I}+\frac{z-1}{3}\left(\mathcal{I}+\mathcal{B}_j^{(\tau)}+(\mathcal{B}_j^{(\tau)})^2\right)\right],\label{uz}\\
	\mathcal{V}(z)
	=\prod_{j=1}^L\left( \mathcal{I}_1^{\otimes(j-1)} \otimes \begin{bmatrix}
			z&1&1\\1&z&1\\1&1&z
	\end{bmatrix} \otimes \mathcal{I}_1^{\otimes(L-j)} \right)
	=\prod_{j=1}^L \left[\left(\mathcal{I}+\mathcal{X}_j+\mathcal{X}_j^2\right) + (z-1)\mathcal{I}\right]\label{vz}
.\end{gather}
Now consider the eigenspaces of $\mathcal{G}=\prod_{j=1}^L \mathcal{X}_j$, denoted as $\mathcal{H}_Q$, where $Q\in\{0,1,-1\}$.
Define the linear map
\begin{equation}\label{mqq}
	\mathcal{M}_{Q\to Q'}\colon\quad \frac{1}{\sqrt{3}}\sum_{s=1}^3 \omega^{-sQ} \left| s, t_2 + s, \cdots, t_L + s \right> \quad\mapsto\quad \frac{1}{\sqrt{3^L}} \sum_{\sigma_1,\cdots,\sigma_L=1}^3 \omega^{\sum_{j=1}^L (t_j-t_{j+1})\sigma_j} \left| \boldsymbol{\sigma} \right>
,\end{equation}
where $t_1\equiv 0$ and $t_{L+1}\equiv Q'$ are fixed, and $t_2,\cdots,t_L\in\{0,1,2\}$.
One can verify that taking all $3^{L-1}$ values of $(t_2,\cdots,t_L)$, the left hand side of Eq.~(\ref{mqq}) forms an orthonormal basis of $\mathcal{H}_Q$, and the right hand side forms an orthonormal basis of $\mathcal{H}_{Q'}$.
Therefore, $\mathcal{M}_{Q\to Q'}$ is an isometry from $\mathcal{H}_Q$ to $\mathcal{H}_{Q'}$.
Next, we will show that $\mathcal{M}_{Q\to \tau}$ connects $\mathcal{T}^{(\tau,Q)}\equiv\mathcal{T}^{(\tau)}|_{\mathcal{H}_Q}$ and $\mathcal{T}^{(Q,\tau)}\equiv\mathcal{T}^{(Q)}|_{\mathcal{H}_\tau}$, and denote $\mathcal{M}_{Q\to \tau}$ as $\mathcal{M}$ for simplicity.
Note that the left hand side of Eq.~(\ref{mqq}) is an eigenvector of $\mathcal{B}_j^{(\tau)}$ while the right hand side is an eigenvector of $\mathcal{X}_j$, and both their eigenvalues equal $\omega^{t_{j+1}-t_j}$.
Hence, we have $\mathcal{M} \mathcal{B}_j^{(\tau)}|_{\mathcal{H}_Q}=\mathcal{X}_j \mathcal{M}$.
Also, note that applying $\mathcal{X}_j$ on the left will multiply each term on the right by $\omega^{\sigma_j-\sigma_{j-1}}$ if $j>1$, and by $\omega^{Q+\sigma_1-\sigma_L}$ if $j=1$.
This is equivalent to applying $\mathcal{B}_{j-1}^{(Q)}$ on the right (define $\mathcal{B}_0^{(Q)}\equiv\mathcal{B}_L^{(Q)}$), implying that $\mathcal{M}\mathcal{X}_j|_{\mathcal{H}_Q} = \mathcal{B}_{j-1}^{(Q)} \mathcal{M}$.
So far, we can see that $\mathcal{M}$ transforms the matrices on $\mathcal{H}_Q$ to matrices on $\mathcal{H}_\tau$ as
\begin{align}
	\mathcal{M}\left(\mathcal{B}_j^{(\tau)}\big|_{\mathcal{H}_Q}\right)\mathcal{M}^{-1} &= \mathcal{X}_j\big|_{\mathcal{H}_\tau},\\
	\mathcal{M}\left(\mathcal{X}_j\big|_{\mathcal{H}_Q}\right)\mathcal{M}^{-1} &= \mathcal{B}_{j-1}^{(Q)}\big|_{\mathcal{H}_\tau}
.\end{align}
Consequently, Eqs.~(\ref{uz}) and (\ref{vz}) are transformed as
\begin{gather}
	\mathcal{M}\left[\mathcal{U}^{(\tau)}(z)\big|_{\mathcal{H}_Q}\right]\mathcal{M}^{-1}= \prod_{j=1}^L \left[\mathcal{I} + \frac{z-1}{3}\left(\mathcal{I} + \mathcal{X}_j + \mathcal{X}_j^2 \right)\right]_{\mathcal{H}_\tau}=\left(\frac{z-1}{3}\right)^L \mathcal{V}(z^\star)\big|_{\mathcal{H}_\tau},\\
	\mathcal{M}\left[\mathcal{V}(z)\big|_{\mathcal{H}_Q}\right]\mathcal{M}^{-1} = \prod_{j=1}^L \left[\left(\mathcal{I} + \mathcal{B}_{j-1}^{(Q)} + (\mathcal{B}_{j-1}^{(Q)})^2\right) + (z-1)\mathcal{I}\right]_{\mathcal{H}_\tau} = (z-1)^L \mathcal{U}^{(Q)}(z^\star ) \big|_{\mathcal{H}_\tau}
.\end{gather}
Together, we have $\mathcal{T}^{(\tau,Q)}(z) = \mathcal{U}^{(\tau)}(z)\mathcal{V}(z)|_{\mathcal{H}_Q}$ similar to
\begin{equation}\label{sim}
\mathcal{M}\left(\mathcal{U}^{(\tau)}(z)\mathcal{V}(z)\big|_{\mathcal{H}_Q}\right)\mathcal{M}^{-1} = 3^{-L}(z-1)^{2L} \mathcal{V}(z^\star )\mathcal{U}^{(Q)}(z^\star ) \big|_{\mathcal{H}_\tau}
.\end{equation}
For any complex $z$, at least one of $\mathcal{V}(z^\star )|_{\mathcal{H}_\tau}$ and $\mathcal{U}^{(Q)}(z^\star )|_{\mathcal{H}_\tau}$ is invertible, so that their product is always similar to $\mathcal{T}^{(Q,\tau)}(z^\star ) = \mathcal{U}^{(Q)}(z^\star )\mathcal{V}(z^\star )|_{\mathcal{H}_\tau}$.